\documentclass[prb,aps,twocolumn]{revtex4-1}

\usepackage{color,amsmath,amssymb,amsfonts,graphicx,tabularx}

\usepackage[unicode=true,colorlinks=true]{hyperref}

\hypersetup{linkcolor=blue,citecolor=blue,urlcolor=blue}

\begin{document}

\title{Non-Abelian Parton Fractional Quantum Hall Effect in Multilayer Graphene}

\author{Ying-Hai Wu$^{1}$, Tao Shi$^{1}$, Jainendra K. Jain$^{2}$}

\affiliation{$^{1}$Max-Planck-Institut f{\"u}r Quantenoptik, Hans-Kopfermann-Stra{\ss}e 1, 85748 Garching, Germany \\
$^{2}$Department of Physics, 104 Davey Lab, Pennsylvania State University, University Park, Pennsylvania 16802, USA}

\begin{abstract}
The current proposals for producing non-Abelian anyons and Majorana particles, which are neither fermions nor bosons, are primarily based on the realization of topological superconductivity in two dimensions. We show theoretically that the unique Landau level structure of bilayer graphene provides a new possible avenue for achieving such exotic particles. Specifically, we demonstrate the feasibility of a ``parton" fractional quantum Hall (FQH) state, which supports non-Abelian particles without the usual topological superconductivity. Furthermore, we advance this state as the fundamental explanation of the puzzling $1/2$ FQH effect observed in bilayer graphene [Kim {\em et al.}, Nano Lett. {\bf 15}, 7445 (2015)], and predict that it will also occur in trilayer graphene. We indicate experimental signatures that differentiate the parton state from other candidate non-Abelian FQH states and predict that a transverse electric field can induce a topological quantum phase transition between two distinct non-Abelian FQH states.
\end{abstract}

\maketitle

The discovery of quantum Hall effect in the early 1980s\cite{Klitzing1980,Tsui1982} ushered in the era of topological phases in modern condensed matter physics. One of the exciting developments it led to was a proposal by Moore and Read\cite{Moore1991,Read2000} who modeled the 5/2 fractional quantum Hall (FQH) effect \cite{Willett1987} as a topological (chiral $p$-wave) superconductor of composite fermions\cite{Jain1989-1}, described by either the so-called Pfaffian wave function\cite{Moore1991} or its hole conjugate called the anti-Pfaffian wave function\cite{Levin2007,LeeSS2007}. They further showed that the vortices of this superconductor bind Majorana zero modes exhibiting non-Abelian braid statistics. The possible application of non-Abelian anyons in topological quantum computation \cite{Kitaev2003,Nayak2008} has inspired intense experimental effort \cite{Radu2008,Willett2009,AnS2011,LinX2012,Willett2013,FuH2016} toward testing the non-Abelian nature of the excitations of the $5/2$ state. The physics of the 5/2 state also served as a paradigm for proposals of topological superconductivity and Majorana modes in other systems  \cite{Alicea2012,Beenakker2013,Elliott2015}.

This article presents the possibility that bilayer graphene can provide a different route to the realization of non-Abelian particles. To date, high mobility GaAs quantum wells have produced the most extensive FQH states. The atomically thin graphene provides another invaluable system for studying quantum Hall physics.  An advantage of FQH states in graphene is its accessibility to direct experimental probes, such as scanning tunneling microscope, which may enable a manipulation of the quasiparticles of FQH states to reveal and perhaps utilize their exotic braid properties. These direct probes are not possible for GaAs quantum wells buried deep below the sample surface. A plethora of FQH states have already been observed in monolayer graphene \cite{DuX2009,Bolotin2009,Feldman2012,Amet2015}, which manifest rich patterns due to the SU(4) spin-valley symmetry, but all of them have odd denominators and can be modeled as integer quantum Hall (IQH) states of composite fermions with spin and valley indices \cite{Balram2015-1}. The absence of FQH effect at half filling in any Landau level (LL) of monolayer graphene has been disappointing but anticipated by theoretical calculations \cite{Toke2007-2,Wojs2011,Balram2015-2}. This situation recently changed dramatically due to the observation of FQH states at even-denominator fractions in {\em bilayer} graphene \cite{KiDK2014,KimYW2015,Zibrov2016,LiJIA2017} (in addition to many other FQH states \cite{Kou2014,Maher2014,Diankov2016}). The FQH states at $\nu=-5/2$, $-1/2$, $3/2$ and $7/2$ most likely corresponds to half-filled $N=1$ LL, and has been proposed \cite{Papic2014} to originate from the Pfaffian or anti-Pfaffian pairing of composite fermions by analogy to the $5/2$ state in GaAs. The physical origin of the $\nu=1/2$ FQH state \cite{KimYW2015}, which nominally corresponds to half-filled $N=0$ LL, has remained a puzzle because one would a priori expect a compressible composite fermion (CF) Fermi liquid state \cite{Halperin1993}.

We demonstrate in this Letter that bilayer graphene can support a new kind of FQH state at $\nu=1/2$ proposed by Jain\cite{Jain1989-2,Jain1990}, called the 221 parton state (the Jain $rst$ parton states defined generally below are to be distinguished from bilayer Halperin $mnl$ states \cite{Halperin1983}). This state also supports fractionally charged excitations with non-Abelian braid statistics \cite{WenXG1998} but does not represent a chiral pairing of composite fermions, and is topologically distinct from the Pfaffian and anti-Pfaffian states. The 221 parton state is not stabilized by any realistic Hamiltonians relevant to semiconductor systems. However, as we show below, the existence of nearly degenerate LLs with different orbital indices in multilayer graphene produces the ideal conditions for generating this state, which we propose to identify with the observed $1/2$ state. A unique feature of the 221 parton state, which sets it apart from all previously observed FQH states, is that it owes its existence fundamentally to LL mixing, disappearing when LL mixing vanishes in the limit of large LL splitting. (LL mixing is believed to break the tie between the Pfaffian and anti-Pfaffian states \cite{Bishara2009,Wojs2010,Simon2013,Peterson2013,Sodemann2013,Peterson2014,Pakrouski2015}, but they occur even without LL mixing.) 

\begin{figure}
\includegraphics[width=0.4\textwidth]{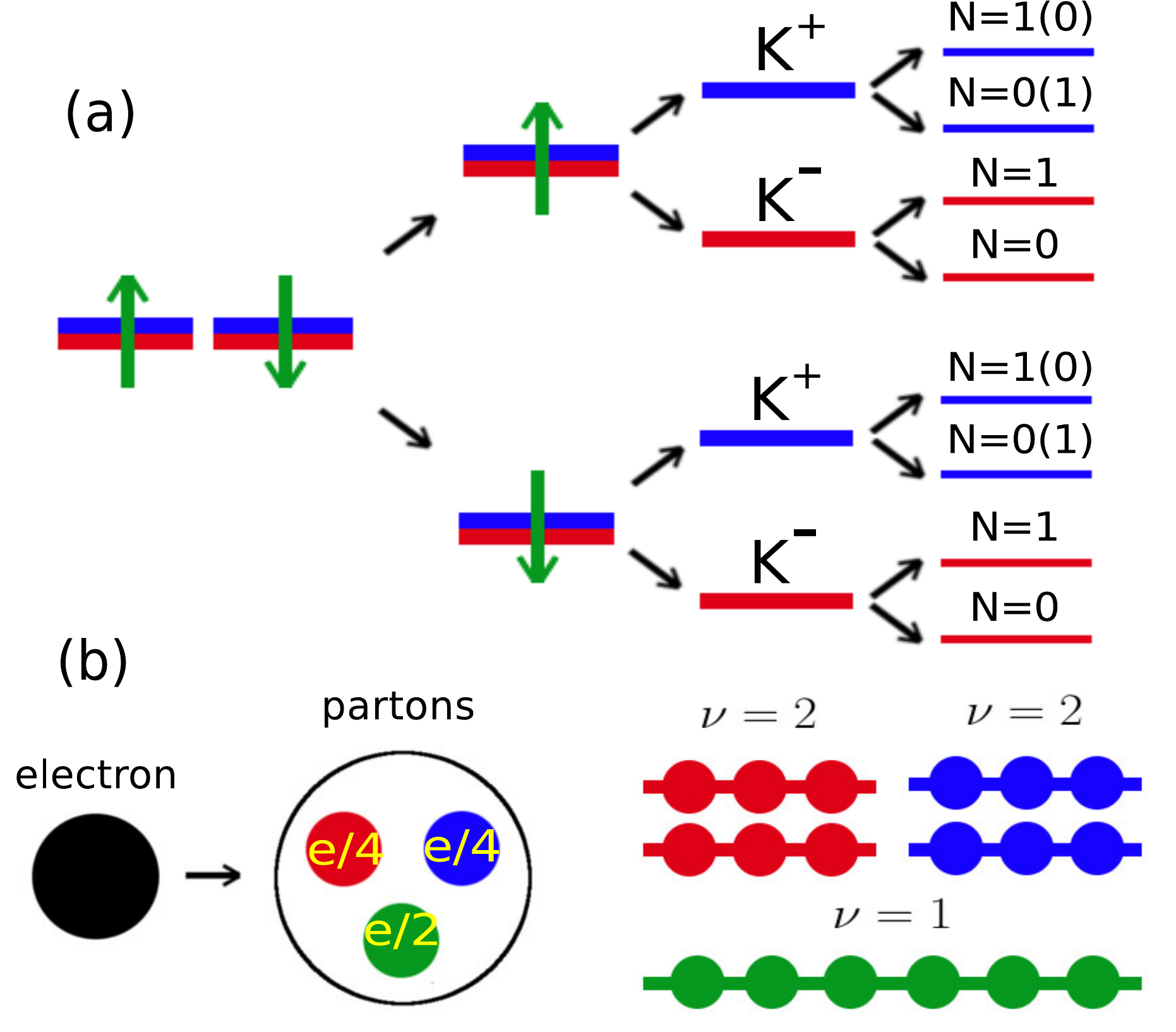}
\caption{(a) The zeroth LL of BLG contains eight non-relativistic LLs because of two orbital indices ($N=0,1$), two spins (green arrows), and two valleys ${\mathbf K}^{\pm}$ (blue and red lines). The degeneracy between these states is found to split as shown \cite{Cote2010,Apalkov2011,Snizhko2012}. The orbital ordering in the ${\mathbf K}^{+}$ valley can be reversed by applying a transverse electric field as shown by the numbers in parentheses. The zero energy LL of TLG has a similar structure where each set contains three non-relativistic LLs with orbital indices $N=0,1,2$. We note that for our purposes the question of whether the spin or the valley symmetry breaking is dominant is not relevant; the 221 parton state only relies on a nearly degenerate doublet of $N=0$ and $N=1$ orbital levels. (b) Schematic of the construction of the 221 parton state. One electron is decomposed into three partons carrying fractional charges, which condense into IQH states with filling factors $\nu=2$, $2$, and $1$.}
\label{Figure1}
\end{figure}

The low-energy physics of Bernel stacked bilayer graphene (BLG) and ABC stacked trilayer graphene (TLG) can be approximately described by chiral fermion models \cite{McCann2006,Barlas2012}. There are two inequivalent valleys ${\mathbf K}^{+}$ and ${\mathbf K}^{-}$ in the Brillouin zone. The coupling to a magnetic field results in the Hamiltonian
\begin{eqnarray}
H_{{\mathbf K}^{+}} = T_J \left[
\begin{array}{cc}
0 & (\pi_x+i\pi_y)^J \\
(\pi_x-i\pi_y)^J & 0
\end{array}
\right]
\label{SingleHamilton}
\end{eqnarray}
for the ${\mathbf K}^{+}$ valley and $H_{{\mathbf K}^{-}}=H^*_{{\mathbf K}^{+}}$ for the ${\mathbf K}^{-}$ valley, where $\pi_i=p_i-eA_i$ is the canonical momentum operator, $J=2$ ($3$) for BLG (TLG) is the chirality, and $T_J$ is a constant depending on microscopic details. The zeroth LL of Eq. (\ref{SingleHamilton}) contains $J$-fold degenerate states $f^{\alpha}_{0}$, $\cdots$, and $f^{\alpha}_{J-1}$ as illustrated in Fig. \ref{Figure1} (a), where $f^{\alpha}_{N}$ are non-relativistic LL states ($N$ is the LL index and $\alpha$ labels the states within each LL). For the disk geometry, we have $f^{\alpha}_{N}(\mathbf{r}) \sim z^{\alpha} L^{\alpha}_{N}(|z|^2/2) e^{-|z|^2/4}$, where $z=(x+iy)/\ell_{B}$ is the holomorphic coordinate and $\ell_{B}=\sqrt{{\hbar}c/(eB)}$ is the magnetic length.

Taking into account the spin and valley degrees of freedom, the $4J$ non-relativistic LLs span the filling factor range $-2J{\leq\nu\leq}2J$. For neutral BLG and TLG at filling factor $\nu=0$, half of these zero energy states are occupied, which we expect to be two subsets with the same spin and valley indices, because this quantum Hall spin-valley ferromagnet \cite{LeeK2014,Datta2016} can efficiently minimize the exchange correlation energy. The FQH states in the interval $0<\nu<J$ are likely to be spin- and valley-polarized so we focus on a set of non-relativistic LLs with orbital indices $N=0\cdots,J-1$. The degeneracy of these LLs is by no means perfect, and the splitting between them can be tuned, e.g. by applying a transverse electric field \cite{KimYW2015,Cote2010,Apalkov2011,Snizhko2012}. We choose below the single-particle Hamiltonian $H_{0}$ to describe electrons in $N=0\cdots,J-1$ non-relativistic LLs separated by a cyclotron energy gap $\hbar\omega_{c}$. In the second quantized notation,
\begin{eqnarray}
H_{0} = \sum^{J-1}_{N=0} \sum_{\alpha} (N+1/2) \hbar\omega_{c} C^{\dagger}_{N\alpha} C_{N\alpha},
\end{eqnarray}
where $C^{\dagger}_{N\alpha}$ ($C_{N\alpha}$) is the creation (annihilation) operator for $f^{\alpha}_{N}$. The interaction term is
\begin{eqnarray}
V = \sum^{J-1}_{\{N_i=0\}} \sum_{\{\alpha_i\}} V^{\{N_i\}}_{\{\alpha_i\}} C^{\dagger}_{N_1\alpha_1} C^{\dagger}_{N_2\alpha_2} C_{N_4\alpha_4} C_{N_3\alpha_3}.
\end{eqnarray}
In our calculations below, we use the Coulomb potential $V_{\rm Coul}({\mathbf r}_j-{\mathbf r}_k)=e^2/(\varepsilon|{\mathbf r}_j-{\mathbf r}_k|)$ ($\varepsilon$ is the dielectric constant of the system) and also a modified interaction with a stronger short-range part. The latter is motivated by the fact that the short-range part of the interaction can be enhanced relatively either due to screening of the Coulomb interaction by interband excitations \cite{Papic2014} or by a dielectric plate on top of the sample \cite{Papic2011}. The coefficients $V^{\{N_i\}}_{\{\alpha_i\}}$ in these cases are given in the Appendix.

In the studies of FQH states, a vital role is played by model wave functions, which clarify the physics and can be tested against exact eigenstates of realistic Hamiltonians in finite-size systems. In the parton construction of FQH states \cite{Jain1989-2,Jain1990}, one divides an electron into fictitious fractionally charged particles called partons, places each species of partons in an IQH state, and then reassembles the partons to obtain an electron state. The number of parton species must be odd to ensure antisymmetry under electron exchange. If the IQH state at filling factor $\nu$ is denoted as $\chi_\nu$, the parton FQH state has the general form $\prod_{j}\chi_{\nu_j}$. Its filling factor is $\nu=[\sum_{j} \nu^{-1}_{j}]^{-1}$ and the charge of the parton with filling factor $\nu_j$ is $e_{j}=e\nu/\nu_{j}$. While all the CF FQH states can be obtained within the parton construction, not all parton FQH states admit interpretation in terms of composite fermions. The state relevant to this work is the $221$ parton state\cite{Jain1989-2,Jain1990}
\begin{eqnarray}
\Psi^{\rm part}_{221}(\{\mathbf r\}) &=& \chi_{2}(\{\mathbf r\}) \chi_{2}(\{\mathbf r\}) \chi_{1}(\{\mathbf r\}) 
\label{ManyWave}
\end{eqnarray}
at $\nu=1/2$ as illustrated in Fig. \ref{Figure1} (b), which lies outside the CF class. The parton construction also suggests a topological field theory for this state \cite{WenXG1998}, which contains an $SU(2)_{2}$ Chern-Simons term and implies that its elementary excitations are Ising type anyons \cite{Witten1989}.

By inspection, we can construct a Hamiltonian for which $\Psi^{\rm part}_{221}(\{\mathbf r\})$ is the exact zero energy ground state. The maximal power of the anti-holomorphic coordinate $\bar{z}$ is two in $\Psi^{\rm part}_{221}(\{\mathbf r\})$, so it has non-zero amplitudes in the $N=0,1,2$ non-relativistic LLs. Furthermore, because it vanishes as $|{\mathbf r}|^3$ when two electrons are brought close to each another, it has zero energy with respect to the Trugman-Kivelson interaction \cite{Trugman1985} $V_{1} = 4\pi\ell^{4}_{B}\sum_{j<k} \nabla^2 \delta^{(2)}(\mathbf{r}_{j} - \mathbf{r}_{k})$ expressed in units of the first Haldane pseudopotential \cite{Haldane1983} in the lowest LL. (The interaction $V_{1}$ may seem somewhat strange, but its matrix elements are well defined.)  $\Psi^{\rm part}_{221}(\{\mathbf r\})$ is thus a zero energy eigenstate for a model in which the $N=0,1,2$ LLs are degenerate at zero energy, all other LLs are sent to infinity, and electrons interact with the $V_{1}$ interaction. One can further show that $\Psi^{\rm part}_{221}(\{\mathbf r\})$ is the unique zero energy state of this model at $\nu=1/2$, because other states at the same filling with zero interaction energy necessarily occupy yet higher LLs and are thus disallowed. This model is accessible to numerical diagonalization studies in the spherical geometry \cite{Haldane1983}, in which $N_{\rm e}$ electrons move on the surface of a sphere with $N_{\phi}$ flux quanta passing through it \cite{WuTT1976,WuTT1977}. The total angular momentum $L$ and its $z$ component $L_{z}$ are good quantum numbers. The incompressible state $\Psi^{\rm part}_{221}(\{\mathbf r\})$ manifests as a uniform $L=0$ state at $N_{\phi}=2N_{\rm e}-5$. For $N_{\rm e}=6,8$ at $N_{\phi}=2N_{\rm e}-5$, we have confirmed that there is a unique zero energy state. If the flux increases to $N_{\phi}=2N_{\rm e}-4$, one expects the additional flux to be accommodated by one of the $\chi_{2}$ factors, which leads to a precise prediction for the number of zero energy states and their quantum numbers; these are in agreement with numerical diagonalization results. Because the addition of one flux to $\chi_{2}$ produces two quasiholes, it follows that the local charge of one quasihole of the 221 parton state is $e^*=e/4$. Our model thus exhibits a $1/2$ FQH effect with $\Psi^{\rm part}_{221}(\{\mathbf r\})$ being the exact ground state.

\begin{figure}
\includegraphics[width=0.4\textwidth]{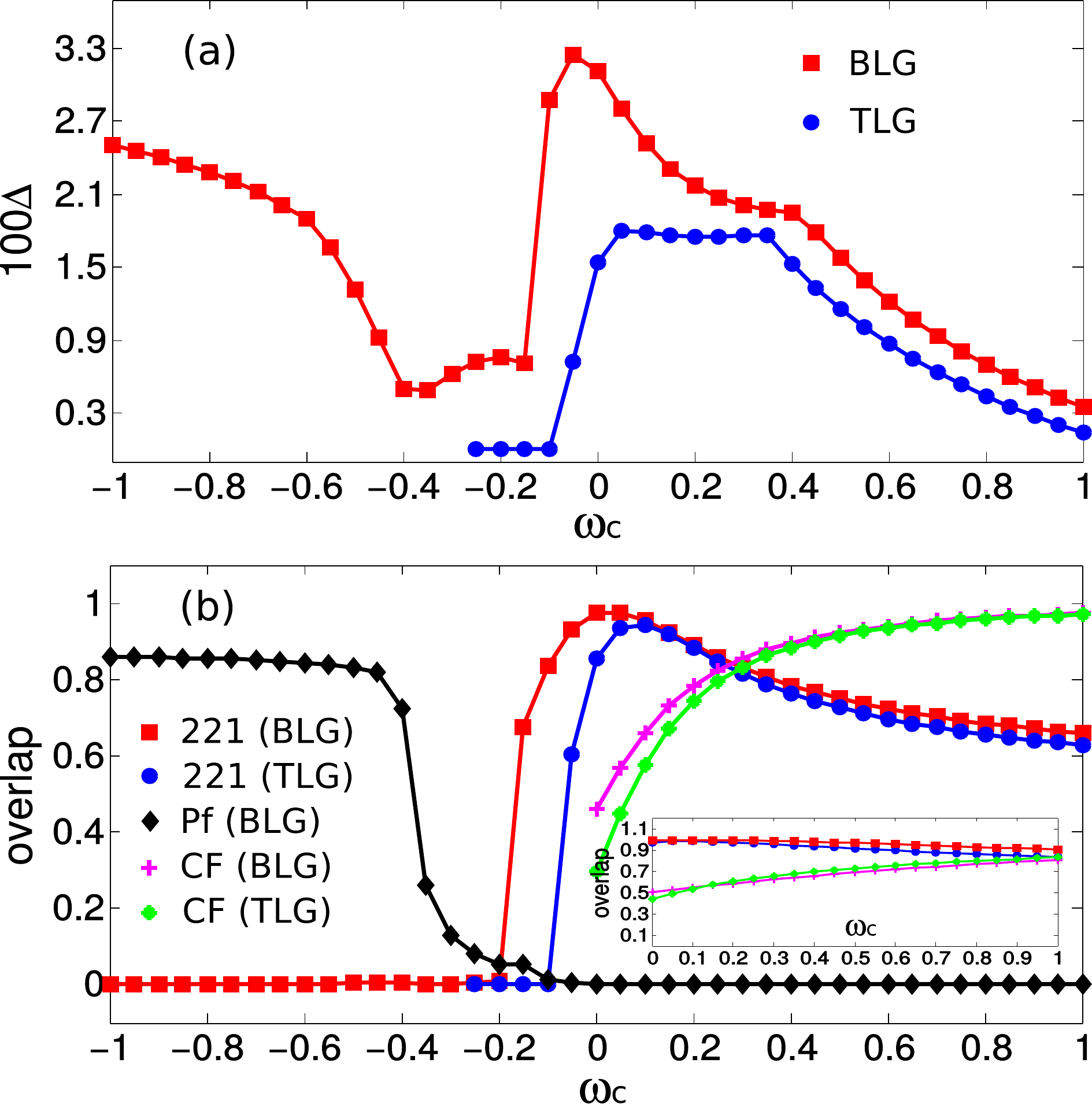}
\caption{Numerical results of the $\nu=1/2$ system with $(N_{\rm e},N_{\phi})=(8,11)$ for a range of LL splitting values $\omega_c$. The Hamiltonian is $H_{0}+V_{\rm Coul}$ and the energy values are quoted in units of $e^2/\epsilon \ell_{\rm B}$. The panel (a) shows the energy gap $\Delta$ of BLG and TLG, where the kinks are due to level crossings of excited states. The panel (b) shows the overlap between the exact ground states of BLG and TLG with various trial wave functions: the 221 parton state, the CF Fermi liquid state, and the Pfaffian state, with symbols indicated in the panel. The inset of (b) shows the overlaps for the Hamiltonian $H_{0}+V_{\rm Coul}+0.2V_{1}$. The $L=0$ subspace contains $418$ ($18212$) states in BLG (TLG).}
\label{Figure2}
\end{figure}

The crucial question is the following: Does this FQH state survive (i.e. the gap does not close) when we vary the interaction from $V_1$ to Coulomb? If so, when is it destabilized as we increase the splitting $\omega_{c}$? To address this, we study the Hamiltonian $H_{0}+V_{\rm Coul}$ numerically. It is customary to restrict the Hilbert space to a single LL when studying FQH states, but we must keep two or three LLs because the $1/2$ state confined to the $N=0$ LL (when $\omega_{c}$ is large) is the CF Fermi liquid state \cite{Halperin1993}. We have studied BLG with $N_{\rm e}=6,8,10,12$ and TLG with $N_{\rm e}=6,8$ at $N_{\phi}=2N_{\rm e}-5$ as a function of $\omega_{c}$. All energies are quoted in units of $e^2/\varepsilon\ell_B$, which is on the order of several hundred Kelvin for typical parameters in graphene. Fig. \ref{Figure2} shows for $(N_{\rm e},N_{\phi})=(8,11)$ the gap $\Delta$ (the energy difference between the ground and the first excited state) as well as the overlap between the exact ground state and $\Psi^{\rm part}_{221}(\{\mathbf r\})$ (projected into the first two LLs for BLG) at various $\omega_{c}$. The high overlaps at small $\omega_c$ ($0.9728$ for BLG and $0.8556$ for TLG at $\omega_c=0$) are very significant in view of the large Hilbert space dimensions (see caption of Fig. \ref{Figure2}). The $(N_{\rm e},N_{\phi})=(10,15)$ system is gapped but aliases with the standard $2/3$ FQH state and is therefore not useful for our purpose. For $(N_{\rm e},N_{\phi})=(12,19)$, we are not able to compute the two lowest energy states in the $L_{z}=0$ subspace due to its large Hilbert space dimension ($\approx2.4\times10^8$), but we have computed the lowest energy states in the $L_{z}=0$ and $1$ subspaces at $\omega_{c}=0$. The energy of the former state is lower by $0.0265$, which tells us that the ground state has $L=0$. If we assume that $0.0265$ is the lowest gap for the $N_{\rm e}=12$ system (i.e. we assume that the lowest excited state has $L{\neq}0$, which is the case for all previously known FQH states) and $\Delta$ has a linear dependence on $1/N_{\rm e}$, the gap at $\omega_{c}=0$ is estimated to be ${\approx}0.017$ in the thermodynamic limit. Fig. \ref{Figure2} also shows the overlap between the exact ground state with the lowest-LL CF state at $L=0$ (a representation of the CF Fermi liquid), which suggests a transition to the CF Fermi liquid at $\omega_c{\sim}0.30$. The inset of Fig. \ref{Figure2} (b) shows the overlap between the exact ground state of the Hamiltonian $H_{0}+V_{\rm Coul}+0.2V_{1}$ and $\Psi^{\rm part}_{221}(\{\mathbf r\})$, which confirms the expectation that when the short range part of the interaction is enhanced (e.g. due to screening \cite{Papic2014}), $\Psi^{\rm part}_{221}(\{\mathbf r\})$ becomes a better approximation and remains valid to larger $\omega_c$. These studies make a strong case for $\Psi^{\rm part}_{221}(\{\mathbf r\})$ at $\nu=1/2$ in BLG and TLG for sufficiently small $\omega_c$. This analysis is also applicable to filling factors $-7/2$, $-3/2$, $5/2$ in BLG and $-11/2$, $-5/2$, $7/2$ in TLG, where one or more sets of the $N=0,\cdots,J-1$ LLs are expected to form a spin-valley ferromagnet and the additional $1/2$ filled states partially occupy one set of $N=0,\cdots,J-1$ LLs as described by our model. We note that Ref. \onlinecite{Papic2014} found weak but inconclusive features at $\nu=-3/2$ in their exact diagonalization study of BLG.

Two other candidates for the $\nu=1/2$ FQH state are the Moore-Read Pfaffian state and its hole partner known as the anti-Pfaffian state \cite{Moore1991,LeeSS2007,Levin2007}, which have been discussed extensively in the context of the $5/2$ FQH state in GaAs \cite{Willett1987}. (There is no ``anti-221" state at $\nu=1/2$ because $\Psi^{\rm part}_{221}(\{\mathbf r\})$ is not confined to a single LL.) The Pfaffian state 
\begin{eqnarray}
\Psi_{\rm Pf}(\{\mathbf r\}) = {\rm Pf} \left( \frac{1}{z_{j}-z_{k}} \right) \prod^{N_{\rm e}}_{j>k=1} \left(z_{j}-z_{k} \right)^2
\end{eqnarray}
occurs at $N_{\phi}=2N_{\rm e}-3$ whereas the anti-Pfaffian state occurs at $N_{\phi}=2N_{\rm e}+1$. The different ``shifts" of Pfaffian, anti-Pfaffian, and 221 parton states indicate their topological distinction. We have also computed the energy spectra of $H_{0}+V_{\rm Coul}$ at $\omega_{c}=0$ for the Pfaffian and anti-Pfaffian shifts. The ground states of $N_{\rm e}=8,10$ in BLG and of $N_{\rm e}=8$ in TLG at $N_{\phi}=2N_{\rm e}-3$ have $L{\neq}0$, which eliminates the Pfaffian state. The ground states of $N_{\rm e}=8$ in BLG and TLG at $N_{\phi}=2N_{\rm e}+1$ have $L=0$, but the energy gaps are only $0.0056$ and $0.0022$ and the overlaps with the anti-Pfaffian state are $0.5820$ and $0.4438$, which suggests that the anti-Pfaffian state is less favored than the 221 parton state. The next system for testing the anti-Pfaffian state at $(N_{\rm e},N_{\phi})=(10,21)$ aliases with the $2/5$ CF state and is thus not useful.

We next predict the exciting possibility of a topological quantum phase transition between two non-Abelian states, namely the 221 and the (anti-)Pfaffian, at filling factors $\nu=-3/2$ and $5/2$ in BLG. This is because the $N=1$ LL can be pushed below the $N=0$ LL when $-2<\nu<0$ and $2<\nu<4$ by applying a transverse electric field \cite{KimYW2015,Cote2010,Apalkov2011,Snizhko2012} as shown in Fig. \ref{Figure1} (a), which means that both positive and negative values of $\omega_{c}$ are physically meaningful at $\nu=-3/2$ and $5/2$. Fig. \ref{Figure2} shows that the overlap between $\Psi^{\rm part}_{221}(\{\mathbf r\})$ and the exact ground state goes down sharply as $\omega_c$ decreases toward the negative side and remains nearly zero ($<0.01$), whereas the energy gap first decreases and then increases. This can be understood as a transition from the 221 parton state into the Pfaffian state, the latter occurring when the electrons predominantly occupy the $N=1$ LL at sufficiently negative $\omega_{c}$. This physics is confirmed by the fact that the overlap between the exact ground state and the Pfaffian state (suitably modified for the $N=1$ LL) becomes quite high as the gap increases in the negative $\omega_{c}$ regime. For completeness, we have also studied TLG in the negative $\omega_{c}$ regime where the $N=2$ LL is at the bottom. Here the 221 parton state transits into a compressible state at $\omega_c\sim-0.10$, which is consistent with the absence of an incompressible state in the half filled $N=2$ LL.

It is interesting to ask if the 221 parton state could also be relevant for the $\nu=-1/2$ state in BLG. This system can be mapped to a filling factor $-1/2-(-2)=3/2$ in the doublet space of $N=0,1$ LLs. If the electrons almost fully occupy the $N=0$ LL, then the $N=1$ LL is nearly half filled and one may expect the physics to be the same as that of the $5/2$ state in GaAs \cite{Papic2014}. However, if one introduces a negative $\omega_{c}$ to attract more electrons into the $N=1$ LL, it is possible for the {\em holes} of this system to form a 221 parton state. We have found that the $(N_{\rm e},N_{\phi})=(18,11)$ system has a $L=0$ ground state and a non-zero gap for $-0.85{\lesssim\omega_{c}\lesssim}-0.20$, but the overlap between the exact ground state and $\Psi^{\rm part}_{221}(\{\mathbf r\})$ of holes is very low ($\lesssim 0.15$), making $\Psi^{\rm part}_{221}(\{\mathbf r\})$ of holes an unlikely candidate for the $\nu=-1/2$ state in BLG.

The Pfaffian, the anti-Pfaffian, and the 221 parton states can in principle be distinguished experimentally. The local charge of the elementary excitations is $e^{*}=e/4$ for all three states. A dimensionless interaction parameter $g$ can be extracted from the temperature dependence of quasiparticle tunneling into the edge states \cite{Radu2008}. Theory predicts $g=1/4$ for the Pfaffian state and $g=1/2$ for the anti-Pfaffian and the 221 parton states \cite{Radu2008,WenXG1993,Bishara2008}. For an ideal edge without reconstruction, the anti-Pfaffian state is expected to have backward-moving edge modes \cite{Levin2007,LeeSS2007} whereas the Pfaffian and the 221 parton states are not, which can be probed in shot noise and local thermometry measurements \cite{Bid2010,Venkat2012}. A combination of these three experiments can help to identify the nature of the experimentally observed state.

The $\nu=1/2$ FQH state was not reported in two recent experiments \cite{Zibrov2016,LiJIA2017}, so it would be important to ascertain the experimental parameters where this state can be observed. Fig. \ref{Figure2} suggests that this state occurs when the $N=1$ level lies at a slightly higher energy $\omega_c$ than the $N=0$ level, i.e. when the LL mixing parameter $(e^2/{\varepsilon}\ell_{B})/(\hbar \omega_c)$ is larger than $\sim$3 ($\omega_c$ can be renormalized by many-body effects \cite{Shizuya2010}). A detailed calculation of the microscopic parameters and the transverse electric field for realizing such conditions is outside the scope of the present work. Interestingly, the optimal parameters at $\nu=-3/2$ and $5/2$ can be determined by studying the crossing transition between $N=0$ and $N=1$ levels as a function of the transverse electric field \cite{Hunt2016}. The parton state is expected to occur close to the transition on the side where the $N=0$ level has a lower energy.

In summary, we have made a convincing case that the 221 parton state should occur for appropriate parameters in BLG and TLG when the ordering of the various LLs is as shown in Fig. \ref{Figure1}. It is natural to associate this state with the observed $\nu=1/2$ FQH state in BLG \cite{KimYW2015} and possibly in TLG, where preliminary evidence for a $\nu=1/2$ FQH state has been reported \cite{BaoWZ2010}. Further experimental and theoretical works will be needed to confirm this identification. In particular, a secure determination of the energy ordering as well as splittings of the various LLs is required (see Refs. \citenum{ShiY2016,Hunt2016} for recent progress in this direction). Looking ahead, multilayer graphene has the potential to host other Jain $rst$ parton states $\Psi^{\rm part}_{rst}(\{\mathbf r\}) = \chi_{r}(\{\mathbf r\}) \chi_{s}(\{\mathbf r\}) \chi_{t}(\{\mathbf r\})$, such as the 331 and 222 states at filling factors $3/5$ and $2/3$. They are distinct from the standard lowest LL states at these filling factors, as they occur at different shifts and possess non-Abelian excitations \cite{WenXG1998}. The realization of these states will open the door to studying topological quantum phase transitions between various FQH states as a function of the LL splitting. The unique LL structure of multilayer graphene thus offers the promise of many new fascinating states and phenomena.

{\em Acknowledgements} --- This work was supported by the EU project {\em Simulations and Interfaces with Quantum Systems} at MPQ and by the U. S. National Science Foundation under grant number DMR-1401636 at Penn State. Exact diagonalization calculations were performed using the DiagHam package for which we are grateful to all the authors. 

\appendix*

\begin{widetext}

\section{Hamiltonian Matrix Elements}

We give the Hamiltonian matrix elements for the cases of our interest. To be consistent with most works in the literature, we define half of the flux through the sphere as $Q=N_{\phi}/2$. The wave functions on sphere are~\cite{WuTT1976}
\begin{eqnarray}
&& Y^{Q}_{l\alpha}(\theta,\phi) = \sqrt{\frac{2l+1}{4\pi}\frac{(l-\alpha)!(l+\alpha)!}{(l-Q)!(l+Q)!}} \;\; \sum^{l-\alpha}_{s=0} (-1)^{l-\alpha+s} \binom{l-Q}{s} \binom{l+Q}{l-\alpha-s} u^{Q+\alpha} v^{Q-\alpha} (u^*u)^s (v^*v)^{l-Q-s}
\end{eqnarray}
where $l=Q+N$ ($N$ is the Landau level index) is the angular momentum, $\alpha$ is the $z$ component of angular momentum, $\theta$ and $\phi$ are the azimuthal and radial angles, and $u=\cos(\theta/2)e^{i\phi/2}$, $v=\sin(\theta/2)e^{-i\phi/2}$ are the spinor coordinates. The magnetic length on sphere is given by $\ell_B = R/\sqrt{Q}$. The monopole harmonics have the properties~\cite{WuTT1977}
\begin{eqnarray}
&& \left[ \psi^{Q}_{l,\alpha} \right]^* = (-1)^{Q-\alpha} \psi^{-Q}_{l,-\alpha} \;\;\;\;\;\; \psi^{Q_1}_{l_1,\alpha_1} \psi^{Q_2}_{l_2,\alpha_2} = \sum^{l_1+l_2}_{L={\rm max}(Q_1+Q_2,|l_1-l_2|)} S_{L} \psi^{Q_1+Q_2}_{L,\alpha_1+\alpha_2} \\
&& S_{L} = (-1)^{3l_1-l_2+L-2Q_1-2Q_2} \sqrt{\frac{(2l_1+1)(2l_2+1)}{4\pi(2L+1)}} \langle l_1,-\alpha_1;l_2,-\alpha_2 | L,-\alpha_1-\alpha_2 \rangle \langle l_1,Q_1;l_2,Q_2 | L,Q_1+Q_2 \rangle
\end{eqnarray}
The matrix elements are
\begin{eqnarray}
V^{\{N_i\}}_{\{\alpha_i\}} = F^{l_1l_2l_4l_3}_{\alpha_1\alpha_2\alpha_4\alpha_3} = \int d{\mathbf \Omega}_1 d{\mathbf \Omega}_2 \; \left[ \psi^{Q}_{l_1\alpha_1}({\mathbf \Omega}_1) \right]^* \left[ \psi^{Q}_{l_2\alpha_2}({\mathbf \Omega}_2) \right]^* V\left( {\mathbf r}_1-{\mathbf r}_2 \right) \psi^{Q}_{l_4\alpha_4}({\mathbf \Omega}_2) \psi^{Q}_{l_3\alpha_3}({\mathbf \Omega}_1)
\end{eqnarray}
where ${\mathbf r}=R(\sin\theta\cos\phi,\sin\theta\sin\phi,\cos\theta)$ and ${\mathbf \Omega}={\mathbf r}/R$. The two identities
\begin{eqnarray}
&& \left[ \psi^{Q}_{l_1\alpha_1}({\mathbf \Omega}_1) \right]^* \left[ \psi^{0}_{LM}({\mathbf \Omega}_1) \right]^* \psi^{Q}_{l_3\alpha_3}({\mathbf \Omega}_1) = \sum^{l_1+l_3}_{L_1=|l_1-l_3|} (-1)^{Q-\alpha_1} S^{1}_{L_1} \left[ \psi^{0}_{LM}({\mathbf \Omega}_1) \right]^* \psi^{0}_{L_1,\alpha_3-\alpha_1}({\mathbf \Omega}_1) \\
&& \left[ \psi^{Q}_{l_2\alpha_2}({\mathbf \Omega}_2) \right]^* \psi^{0}_{LM}({\mathbf \Omega}_2) \psi^{Q}_{l_4\alpha_4}({\mathbf \Omega}_2) = \sum^{l_2+l_4}_{L_2=|l_2-l_4|} (-1)^{Q-\alpha_4} S^{2}_{L_2} \left[ \psi^{0}_{L_2,\alpha_2-\alpha_4}({\mathbf \Omega}_2) \right]^* \psi^{0}_{LM}({\mathbf \Omega}_2)
\end{eqnarray}
will be used later for computing $F^{l_1l_2l_4l_3}_{\alpha_1\alpha_2\alpha_4\alpha_3}$. For the short-range interaction $V\left( {\mathbf r}_1-{\mathbf r}_2 \right)=4\pi\ell^4_{B}\nabla^2\delta\left( {\mathbf r}_1-{\mathbf r}_2 \right)$, the matrix elements are
\begin{eqnarray}
&& \sum^{\infty}_{L=0} \sum^{L}_{M=-L} - \frac{4\pi}{Q^2} L(L+1) \int d{\mathbf \Omega}_1 d{\mathbf \Omega}_2 \; \left[ \psi^{Q}_{l_1\alpha_1}({\mathbf \Omega}_1) \right]^* \left[ \psi^{0}_{LM}({\mathbf \Omega}_1) \right]^* \psi^{Q}_{l_3\alpha_3}({\mathbf \Omega}_1) \left[ \psi^{Q}_{l_2\alpha_2}({\mathbf \Omega}_2) \right]^* \psi^{0}_{LM}({\mathbf \Omega}_2) \psi^{Q}_{l_4\alpha_4}({\mathbf \Omega}_2) \nonumber \\
&& = \delta_{\alpha_1+\alpha_2,\alpha_3+\alpha_4} \sum^{{\rm min}(l_1+l_3,l_2+l_4)}_{L={\rm max}(|l_1-l_3|,|l_2-l_4|)} \frac{4\pi}{Q^2} L(L+1) \; (-1)^{2Q-\alpha_1-\alpha_4+1} S^{1}_{L} S^{2}_{L}
\end{eqnarray}
For the Coulomb interaction $V\left( {\mathbf r}_1-{\mathbf r}_2 \right)=e^2/(\varepsilon|{\mathbf r}_1-{\mathbf r}_2|)$, the relation
\begin{eqnarray}
\frac{1}{|{\mathbf r}_1-{\mathbf r}_2|} = \frac{4\pi}{R} \sum^{\infty}_{L=0} \sum^{L}_{M=-L} \frac{1}{2L+1} \left[ \psi^{0}_{LM}({\mathbf \Omega}_1) \right]^* \psi^{0}_{LM}({\mathbf \Omega}_2)
\end{eqnarray}
helps us to obtain the matrix elements [in units of $e^2/(\varepsilon\ell_{B})$]
\begin{eqnarray}
&& \sum^{\infty}_{L=0} \sum^{L}_{M=-L} \frac{4{\pi}}{\sqrt{Q}(2L+1)} \int d{\mathbf \Omega}_1 d{\mathbf \Omega}_2 \; \left[ \psi^{Q}_{l_1\alpha_1}({\mathbf \Omega}_1) \right]^* \left[ \psi^{0}_{LM}({\mathbf \Omega}_1) \right]^* \psi^{Q}_{l_3\alpha_3}({\mathbf \Omega}_1) \left[ \psi^{Q}_{l_2\alpha_2}({\mathbf \Omega}_2) \right]^* \psi^{0}_{LM}({\mathbf \Omega}_2) \psi^{Q}_{l_4\alpha_4}({\mathbf \Omega}_2) \nonumber \\
&& = \delta_{\alpha_1+\alpha_2,\alpha_3+\alpha_4} \sum^{{\rm min}(l_1+l_3,l_2+l_4)}_{L={\rm max}(|l_1-l_3|,|l_2-l_4|)} \frac{4\pi}{\sqrt{Q}(2L+1)} (-1)^{2Q-\alpha_1-\alpha_4} S^{1}_{L} S^{2}_{L}
\end{eqnarray}

\end{widetext}

\bibliography{../ReferCollect}

\end{document}